# Confirmation Bias and the Open Access Advantage: Some Methodological Suggestions for the Davis Citation Study

Stevan Harnad

Chaire de recherche du Canada en sciences cognitives
Institut des sciences cognitives
Université du Québec à Montréal
Montréal, Québec, Canada H3C 3P8

Department of Electronics and Computer Science
University of Southampton
Highfield, Southampton, United Kingdon SO17 1BJ

**SUMMARY:** *Davis (2008) analyzes citations from 2004-2007 in 11 biomedical journals. For 1,600 of the 11,000 articles (15%), their authors paid the publisher to make them Open Access (OA). The outcome, confirming previous studies (on both paid and unpaid OA), is a significant OA citation Advantage, but a small one (21%, 4% of it correlated with other article variables such as number of authors, references and pages). The author infers that the size of the OA advantage in this biomedical sample has been shrinking annually from 2004-2007, but the data suggest the opposite. In order to draw valid conclusions from these data, the following five further analyses are necessary:*

***(1)*** *The current analysis is based only on author-choice (paid) OA. Free OA self-archiving needs to be taken into account too, for the same journals and years, rather than being counted as non-OA, as in the current analysis.*

***(2)*** *The proportion of OA articles per journal per year needs to be reported and taken into account.*

***(3)*** *Estimates of journal and article quality and citability in the form of the Journal Impact Factor and the relation between the size of the OA Advantage and journal as well as article "citation-bracket" need to be taken into account.*

***(4)*** *The sample-size for the highest-impact, largest-sample journal analyzed, PNAS, is restricted and is excluded from some of the analyses. An analysis of the full PNAS dataset is needed, for the entire 2004-2007 period.*

***(5)*** *The analysis of the interaction between OA and time, 2004-2007, is based on retrospective data from a June 2008 total cumulative citation count. The analysis needs to be redone taking into account the dates of both the cited articles and the citing articles, otherwise article-age effects and any other real-time effects from 2004-2008 are confounded.*

*Davis proposes that an author self-selection bias for providing OA to higher-quality articles (the Quality Bias, QB) is the primary cause of the observed OA Advantage,*

*but this study does not test or show anything at all about the causal role of QB (or of any of the other potential causal factors, such as Accessibility Advantage, AA, Competitive Advantage, CA, Download Advantage, DA, Early Advantage, EA, and Quality Advantage, QA). The author also suggests that paid OA is not worth the cost, per extra citation. This is probably true, but with OA self-archiving, both the OA and the extra citations are free.*

> **Comments on:** Davis, P.M. (2008) Author-choice open access publishing in the biological and medical literature: a citation analysis. *Journal of the American Society for Information Science and Technology* (*JASIST*) (in press) http://arxiv.org/pdf/0808.2428v1

The Davis (2008) preprint is an analysis of the citations from years c. 2004-2007 in 11 biomedical journals: c. 11,000 articles, of which c. 1,600 (15%) were made Open Access (OA) through "Author Choice" (AC-OA): author chooses to pay publisher for OA). Author self-archiving (SA-OA) articles from the same journals was not measured.

The result was a significant OA citation advantage (21%) over time, of which 4% was correlated with variables other than OA and time (number of authors, pages, references; whether article is a Review and has a US co-author).

This outcome confirms the findings of numerous previous studies (some of them based on far larger samples of fields, journals, articles and time-intervals) reporting an OA citation advantage (ranging from 25%-250%) in all fields, across a 10-year range (Hitchcock 2008).

The preprint also states that the size of the OA advantage in this biomedical sample diminishes annually from 2004-2007. But the data seem to show the opposite: that as an article gets older, and its cumulative citations grow, its absolute and relative OA advantage grow too.

The preprint concludes, based on its estimate of the size of the OA citation Advantage, that AC-OA is not worth the cost, per extra citation. This is probably true -- but with SA-OA the OA and the extra citations can be had at no cost at all.

The paper is accepted for publication in *JASIST*. It is not clear whether the linked text is the unrefereed preprint, or the refereed, revised postprint. On the assumption that it is the unrefereed preprint, what follows is an extended peer commentary with recommendations on what should be done in revising it for publication.

(*It is very possible, however, that some or all of these revisions were also recommended by the JASIST referees and that some of the changes have already been made in the published version.*)

As it stands currently, this study (i) confirms a significant OA citation Advantage, (ii) shows that it grows cumulatively with article age and (iii) shows that it is correlated with several other variables that are correlated with citation counts.

Although the author argues that an author self-selection bias for preferentially providing OA to higher-quality articles (the Quality Bias, QB) is the primary causal factor underlying the observed OA Advantage, in fact this study does not test or show anything at all about the causal role of QB (or of any of the other potential causal factors underlying the OA Advantage, such as Accessibility Advantage, AA, Competitive Advantage, CA, Download Advantage, DA, Early Advantage, EA, and Quality Advantage, QA; Hajjem & Harnad 2007b).

The following 5 further analyses of the data are necessary. The size and pattern of the observed results, as well as their interpretations, could all be significantly altered (as well as deepened) by their outcome:

> **(1)** The current analysis is based only on author-choice (paid) OA. Free author self-archiving OA needs to be taken into account too, for the same journals and years, rather than being counted as non-OA, as in the current analysis.
>
> **(2)** The proportion of OA articles per journal per year needs to be reported and taken into account.
>
> **(3)** Estimates of journal and article quality and citability in the form of the Journal Impact Factor (journal's average citations) and the relation between the size of the OA Advantage and journal and article "citation-bracket" need to be taken into account.
>
> **(4)** The sample-size for the highest-impact, largest-sample journal, PNAS, is restricted and is excluded from some of the analyses. A full analysis of the full PNAS dataset is needed, for the entire 2004-2007 period.
>
> **(5)** The analysis of the interaction between OA and time, 2004-2007, is based on retrospective data from a June 2008 total cumulative citation count. The analysis needs to be redone taking into account the dates of both the cited articles and the citing articles, otherwise article-age effects and any other real-time effects from 2004-2008 are confounded.

**Commentary on the text of the preprint:**

> **"ABSTRACT…** ***there is strong evidence to suggest that the open access advantage is declining by about 7% per year, from 32% in 2004 to 11% in 2007"***

It is not clearly explained how these figures and their interpretation are derived, nor is it reported how many OA articles there were in each of these years. The figures appear to be based on a statistical interaction between OA and article-age in a multiple regression analysis for 9 of the 11 journals in the sample. (a) The data from *PNAS*, the largest and highest-impact journal, are excluded from this analysis. (b) The many variables included in the (full) multiple regression equation (across journals) omit one of the most obvious ones: journal impact factor. (c) OA articles that are self-archived rather than paid author-choice are not identified and included as OA, hence their citations are counted as being non-OA. (d) The OA/age interaction is not based on yearly citations after a fixed interval for each year, but on cumulative retrospective citations

in June 2008.

The natural interpretation of Figure 1 accordingly seems to be the exact opposite of the one the author makes: Not that the size of the OA Advantage *shrinks* from 2004-2007, but that the size of the OA Advantage *grows* from 2007-2004 (as articles get older and their citations grow). Not only do cumulative citations grow for both OA and non-OA articles from year 2007 articles to year 2004 articles, but the cumulative OA advantage *increases* (by about 7% per year, even on the basis of this study's rather slim and selective data and analyses).

This is quite natural, as not only do citations grow with time, but the OA Advantage -- barely detectable in the first year, being then based on the smallest sample and the fewest citations -- emerges with time.

> ***"See Craig et al. [2007] for a critical review of the literature [on the OA citation advantage]"***

Craig et al's rather slanted 2007 review is the only reference to previous findings on the OA Advantage cited by the Davis preprint (Harnad 2007a). Craig et al. had attempted to reinterpret the many times replicated positive finding of an OA citation advantage on the basis of 4 negative findings (Davis & Fromerth, 2007; Kurtz et al., 2005; Kurtz & Henneken, 2007; Moed, 2007), in maths, astronomy and condensed matter physics, respectively. Apart from Davis's own prior study, these studies were based mainly on *preprints* that were made OA well before publication. The observed OA advantage consisted mostly of an Early Access Advantage for the OA prepublication preprint, plus an inferred Quality Bias (QB) on the part of authors towards preferentially providing OA to higher quality preprints (Harnad 2007b).

The Davis preprint does not cite any of the considerably larger number of studies that have reported large and consistent OA advantages for *postprints*, based on many more fields, some of them based on far larger samples and longer time intervals (Hitchcock 2008). Instead, Davis focuses rather single-mindedly on the hypothesis that most or all of the OA Advantage is the result for the self-selection bias (QB) toward preferentially making higher-quality (hence more citeable) articles OA:

> ***"authors selectively choose which articles to promote freely… [and] highly cited authors disproportionately choose open access venues"***

It is undoubtedly true that better authors are more likely to make their articles OA, and that authors in general are more likely to make their better articles OA. This *Quality* or "Self-Selection" *Bias* (QB) is one of the probable causes of the OA Advantage.

However, no study has shown that QB is the *only* cause of the OA Advantage, nor even that it is the *biggest* cause. Three of the studies cited (Kurtz et al., 2005; Kurtz & Henneken, 2007; Moed, 2007) showed that another causal factor is *Early Access* (EA: providing OA earlier results in more citations).

There are several other candidate causal factors in the OA Advantage, besides QB and EA

(Hajjem & Harnad 2007b):

| | |
|---|---|
| **QA**<br>**Quality Advantage** | **-- top 20% of articles receive 80% of citations, hence gain more from OA**<br>**-- advantage still present after universal OA**<br>**-- measurable compared to pre-OA** |
| *QB*<br>*Quality Bias (Self-Selection)* | *-- authors more likely to make top articles OA*<br>*-- advantage vanishes with universal OA* |
| **AA**<br>**Accessibility Advantage** | **-- nonsubscriber access**<br>**-- advantage still present after universal OA**<br>**-- measurable compared to pre-OA**<br>**-- higher-quality articles may benefit more** |
| *CA*<br>*Competitive Advantage* | *-- relative advantage*<br>*-- advantage vanishes with universal OA*<br>*-- higher-quality articles may benefit more* |
| **DA**<br>**Download Advantage** | **-- increased downloads (correlated with later citation increase)**<br>**-- advantage still present after universal OA**<br>**-- measurable compared to pre-OA**<br>**-- higher-quality articles may benefit more** |
| **EA**<br>**Early Access Advantage** | **-- earlier access increases downloads and citations**<br>**-- advantage still present after universal OA**<br>**-- measurable compared to pre-OA, and for articles made OA before publication as preprints**<br>**-- higher-quality articles may benefit more** |

There is the *Download* (or Usage) *Advantage* (DA): OA articles are downloaded significantly more, and this early DA has also been shown to be predictive of a later citation advantage in Physics (Brody et al. 2006).

There is a *Competitive Advantage* (CA): OA articles are in competition with non-OA articles, and to the extent that OA articles are relatively more accessible than non-OA articles, they can be used and cited more. Both QB and CA, however, are *temporary* components of the OA advantage that will necessarily shrink to zero and disappear once all research is OA. EA and DA, in contrast, will continue to contribute to the OA advantage even after universal OA is reached, when all postprints are being made OA immediately upon publication, compared to pre-OA days (as Kurtz has shown for Astronomy, which has already reached universal post-publication OA).

There is an *Accessibility Advantage* (AA) for those users whose institutions do not have subscription access to the journal in which the article appeared. AA too (unlike CA) persists even after universal OA is reached: all articles then have AA's full benefit.

And there is at least one more important causal component in the OA Advantage, apart from AA, CA, DA and QB, and that is a *Quality Advantage* (QA), which has often been erroneously conflated with QB (Quality Bias):

Ever since Lawrence's original study in 2001, the OA Advantage can be estimated in two different ways: (1) by comparing the average citations for OA and non-OA articles (log citation ratios *within the same journal and year,* or regression analyses like Davis's) and (2) by comparing the proportion of OA articles in different "citation brackets" (0, 1, 2, 3-4, 5-8, 9-16, 17+ citations).

In method (2), the OA Advantage is observed in the form of an increase in the proportion of OA articles in the higher citation brackets. But this correlation can be explained in two ways. One is QB, which is that authors are more likely to make higher-quality articles OA. But it is also at least as plausible that *higher-quality articles benefit more from OA*! It is already known that the top c. 10-20% of articles receive c. 80-90% of all citations (Seglen's 1992 "skewness of science"). It stands to reason, then, that when all articles are made OA, it is the top 20% of articles that are most likely to be cited more: Not all OA articles benefit from OA equally, because not all articles are of equally citable quality.

Hence both QB and QA are likely to be causal components in the OA Advantage, and the only way to tease them apart and estimate their individual contributions is to control for the QB effect by *imposing* the OA instead of allowing it to be determined by self-selection. We (Gargouri, Hajjem, Gingras, Carr & Harnad, in prep.) are completing such a study now, comparing mandated and unmandated OA; and Davis et al 2008 have just published another study on randomized OA for 11 journals:

> ***"In the first controlled trial of open access publishing where articles were randomly assigned to either open access or subscription-access status, we recently reported that no citation advantage could be attributed to access status (Davis, Lewenstein, Simon, Booth, & Connolly, 2008)"***

This randomized OA study by Davis et al. was very welcome and timely, but it had originally been announced to cover a 4-year period, from 2007-2010, whereas it was instead prematurely published in 2008, after only one year. No OA advantage at all was observed in that 1-year interval, and this too agrees with the many existing studies on the OA Advantage, some based on far larger samples of journals, articles and fields: Most of those studies (none of them randomized) likewise detected no OA citation advantage at all in the first year: It is simply too early. In most fields, citations take longer than a year to be made, published, ISI-indexed and measured, and to make any further differentials (such as the OA Advantage) measurable. (This is evident in Davis's present preprint too, where the OA advantage is barely visible in the first year (2007).)

The only way the *absence* of a significant OA advantage in a sample with randomized OA can be used to demonstrate that the OA Advantage is only or mostly just a self-selection bias (QB) is by also demonstrating the *presence* of a significant OA advantage in the same (or comparable) sample with *nonrandomized* (i.e., self-selected) OA.

But Davis et al. did not do this control comparison (Harnad 2008b). Finding no OA Advantage with randomized OA after one year merely confirms the (widely observed) finding that one year is usually too early to detect any OA Advantage; but it shows nothing whatsoever about self-selection QB.

> *"we examine the citation performance of author-choice open access"*

It is quite useful and interesting to examine citations for OA and non-OA articles where the OA is provided through (self-selected) "Author-Choice" (i.e., authors paying the publisher to make the article OA on the publisher's website).

Most prior studies of the OA citation Advantage, however, are based on *free self-archiving by authors on their personal, institutional or central websites*. In the bigger studies, a robot trawls the web using ISI bibliographic metadata to find which articles are freely available on the web (Hajjem et al. 2005).

Hence a natural (indeed essential) control test that has been omitted from Davis's current author-choice study – a test very much like the control test omitted from the Davis et al randomized OA study – is to *identify the articles in the same sample that were made OA through author self-archiving*. If those articles are identified and counted, that not only provides an estimate of the relative uptake of author-choice OA vs OA self-archiving in the same sample interval, but it allows a comparison of their respective OA Advantages. More important, it corrects the estimate of an OA Advantage based on author-choice OA alone: For, as Davis has currently done the analysis, any OA Advantage from OA self-archiving in this sample would in fact *reduce* the estimate of the OA Advantage based on author-choice OA (mistakenly counting as non-OA the articles and citation-counts for self-archived OA articles)

> ***"METHODS… The uptake of the open access author-choice programs for these [11] journals ranged from 5% to 22% over the dates analyzed"***

Davis's preprint does not seem to provide the data – either for individual journals or for the combined totals – on the percentage of author-choice OA (henceforth AC-OA) by year, nor on *the relation between the proportion uptake of AC-OA and the size of the OA Advantage, by year*.

As Davis has been careful to do multiple regression analyses on many of the article-variables that might correlate with citations and OA (article age, number of authors, number of references, etc.), it seems odd not to take into account the relation between the size of the AC-OA Advantage and the degree of uptake of AC-OA, by year. The other missing information is the corresponding data for self-archiving OA (henceforth SA-OA).

> ***"[For] All of the journals… all articles roll into free access after an initial period [restricted to subscription access only for 12 months (8 journals), 6 months (2 journals) or 24 months (1 journal)]"***

(This is important in relation to the Early Access (EA) Advantage, which is the biggest

contributor to the OA Advantage in the two cited studies by Kurtz on Astronomy. *Astronomy has free access to the postprints of all articles in all astronomy journals immediately upon publication*. Hence Astronomy has scope for an OA Advantage *only* through an EA Advantage, arising from the early posting of preprints before publication. The size of the OA Advantage in other fields -- in which (unlike in Astronomy) access to the postprint is restricted to subscribers-only for 6, 12, or 24 months -- would then be the equivalent of an estimate of an "EA Advantage" for those potential users who lack subscription access – i.e., the Accessibility Advantage, AA.)

> ***"Cumulative article citations were retrieved on June 1, 2008. The age of the articles ranged from 18 to 57 months"***

Most of the 11 journals were sampled till December 2007. That would mean that *the 2007 OA Advantage was based on even less than one year from publication*.

> ***"STATISTICAL ANALYSIS… Because citation distributions are known to be heavily skewed (Seglen, 1992) and because some of the articles were not yet cited in our dataset, we followed the common practice of adding one citation to every article and then taking the natural log"***

(How well did that correct the skewness? If it still was not normal, then citations might have to be dichotomized as a 0/1 variable, comparing, by citation-bracket *slices*, (1) 0 citations vs 1 or more citations, (2) 0 or 1 vs more than 1, (3) 2 or fewer vs. more than 2, (4) 3 or fewer vs. more than 3… etc.)

> ***"For each journal, we ran a reduced [2 predictor] model [article age and OA] and a full [7 predictor] regression model [age, OA; log no. of authors, references, pages; Review; US author]"***

Both analyses are, of course, a good idea to do, but why was *Journal Impact Factor (JIF)* not tested as one of the predictor variables in the cross-journal analyses (Hajjem & Harnad 2007a)? Surely JIF, too, correlates with citations: Indeed, the Davis study assumes as much, as it later uses JIF as the multiplier factor in calculating the cost per extra citation for author-choice OA (see below).

Analyses by journal JIF citation-bracket, for example, can provide estimates of QA (Quality Advantage) if the OA Advantage is bigger in the higher journal citation-brackets. (Davis's study is preoccupied with the self-selection QB bias, which it does not and cannot test, but it fails to test other candidate contributors to the OA Advantage that it *can* test.)

(An important and often overlooked logical point should also be noted about the correlates of citations and the direction of causation: The many predictor variables in the multiple regression equations predict not only the OA citation Advantage; they also predict citation counts themselves. *It does not necessarily follow from the fact that, say, longer articles are more likely to be cited that article length is therefore an artifact that must be factored out of citation counts* *in order to get a more valid estimate of how accurately citations measure quality*. One possibility is that length is indeed an artifact. But the other possibility is that length is a valid causal factor in quality! *If* length is indeed an artifact, then longer articles are being cited more just because they are longer, rather than because they are better, and this length bias needs to be subtracted out of citation counts as measures of quality. But *if the extra length is a causal contributor to what makes the better articles better, then subtracting out the length effect simple serves to make citation counts a blunter, not a sharper instrument for measuring quality*. The same reasoning applies to some of the other correlates of citation counts, as well as their relation to the OA citation Advantage. Systematically removing them all, even when they are not artifactual, systematically divests citation counts of their potential power to predict quality. This is another reason why citation counts need to be systematically validated against other evaluative measures [Harnad 2008a].)

> ***"Because we may lack the statistical power to detect small significant differences for individual journals, we also analyze our data on an aggregate level"***

It is a reasonable, valid strategy, to analyze across journals. Yet this study still persists in drawing individual-journal level conclusions, despite having indicated (correctly) that its sample may be too small to have the power to detect individual-journal level differences (see below).

(On the other hand, it is not clear whether all the OA/non-OA citation comparisons were always *within-journal, within-year*, as they ought to be; no data are presented for the percentage of OA articles per year, per journal. *OA/non-OA comparisons must always be within-journal/year comparisons, to be sure to compare like with like.*)

> ***"The first model includes all 11 journals, and the second omits the Proceedings of the National Academy of Sciences (PNAS), considering that it contributed nearly one-third (32%) of all articles in our dataset"***

Is this a justification for excluding PNAS? Not only was the analysis done with and without PNAS, but, unlike all the other journals, whose data were all included, for the entire time-span, PNAS data were only included from the first and last six months.

Why? PNAS is a very high impact factor journal, with highly cited articles. A study of PNAS alone, with its much bigger sample size, would be instructive in itself – and would almost certainly yield a bigger OA Advantage than the one derived from averaging across all 11 journals (and reducing the PNAS sample size, or excluding PNAS altogether).

There can be a QB difference *between* PNAS and non-PNAS articles (and authors), to be sure, because PNAS publishes articles of higher quality. But a *within*-PNAS year-by-year comparison of OA and non-OA that yielded a bigger OA Advantage than a within-journal OA/non-OA comparison for lower-quality journals would also reflect the contribution of QA. (With these data in hand, the author should not be so focused on confirming his hypotheses: take the opportunity to falsify them too!)

> ***"we are able to control for variables that are well-known to predict future citations [but] we cannot control for the quality of an article"***

This is correct. One cannot control for the quality of an article; but in comparing *within* a journal/year, one can compare the size of the OA Advantage for higher and lower impact journals; if the advantage is higher for higher-impact journals, that favors QA over QB.

One can also take target OA and non-OA articles (within each citation bracket), and match the title words of each target article with other articles (in the same journal/year):

If one examines N-citation OA articles and N-citation non-OA articles, are their title-word-matched (non-OA) control articles equally likely to have N or more citations? Or are the word-matched control articles for N-citation OA articles less likely to have N or more citations than the controls for N-citation non-OA articles (which would imply that the OA has raised the OA article's citation bracket)? And would this effect be greater in the higher citation brackets than in the lower ones (N = 1 to N = >16)?

If one is resourceful, there are ways to control, or at least triangulate on quality indirectly.

> ***"spending a fee to make one's article freely available from a publisher's website may indicate there is something qualitatively different [about that article]"***

Yes, but one could probably tell a Just-So story either way about the *direction* of that difference: paying for OA because one thinks one's article is better, or paying for OA because one thinks one's article is worse! Moreover, this is AC-OA, which costs money; the stakes are different with SA-OA, which only costs a few keystrokes. But this analysis omitted to identify or measure SA-OA.

> **"RESULTS…The difference in citations between open access and subscription-based articles is small and non-significant for the majority of the journals under investigation"**

(1) Compare the above with what is stated earlier: "Because we may lack the statistical power to detect small significant differences for individual journals, we also analyze our data on an aggregate level."

(2) Davis found an OA Advantage across the entire sample of 11 journals, whereas the individual journal samples were too small. Why state this as if it were some sort of an empirical effect?

> ***"where only time and open access status are the model predictors, five of the eleven journals show positive and significant open access effects."***

(That does not sound too bad, considering that the individual journal samples were small and hence lacked the statistical power to detect small significant differences, and that the PNAS sample was made deliberately small!)

> ***"Analyzing all journals together, we report a small but significant increase in article citations of 21%."***

Whether that OA Advantage is small or big remains to be seen. The bigger published OA Advantages have been reported on the basis of bigger samples.

> ***"Much of this citation increase can be explained by the influence of one journal, PNAS. When this journal is removed from the analysis, the citation difference reduces to 14%."***

This reasoning can appeal only if one has a confirmation bias: PNAS is also the journal with the biggest sample (of which only a fraction was used); and it is also the highest impact journal of the 11 sampled, hence the most likely to show benefits from a Quality Advantage (QA) that generates more citations for higher citation-bracket articles. If the objective had not been to demonstrate that there is little or no OA Advantage (and that what little there is is just due to QB), PNAS would have been analyzed more closely and fully, rather than being minimized and excluded.

> ***"When other explanatory predictors of citations (number of authors, pages, section, etc.) are included in the full model, only two of the eleven journals show positive and significant open access effects. Analyzing all journals together, we estimate a 17% citation advantage, which reduces to 11% if we exclude PNAS."***

In other words partialling out 5 more correlated variables from this sample reduces the residual OA Advantage by 4%. And excluding the biggest, highest-quality journal's data, reduces it still further.

If there were not this strong confirmation bent on the author's part, the data would be treated in a rather different way: The fact that a journal with a bigger sample enhances the OA Advantage would be treated as a plus rather than a minus, suggesting that still bigger samples might have the power to detect still bigger OA Advantages. And the fact that PNAS is a higher quality journal would also be the basis for looking more closely at the role of the Quality Advantage (QA). (With less of a confirmation bent, OA Self-archiving, too, would have been controlled for, instead of being credited to non-OA.)

Instead, the awkward persistence of a significant OA Advantage even after partialling out the effects of so many correlated variables, despite restricting the size of the PNAS sample, and even after removing PNAS entirely from the analysis, has to be further explained away:

> ***"The modest citation advantage for author-choice open access articles also appears to weaken over time. Figure 1 plots the predicted number of citations for the average article in our dataset. This difference is most pronounced for articles published in 2004 (a 32% advantage), and***

> ***decreases by about 7% per year (Supplementary Table S2) until 2007 where we estimate only an 11% citation advantage."***

(The methodology is not clearly described. We are not shown the percent OA per journal per year, nor what the dates of the citing articles were, for each cited-article year. What is certain is that a 1-year-old 2007 article differs from a 4-year-old 2004 article not just in its total cumulative citations in June 2008, but in that the estimate of its citations per year is based on a much smaller sample, again reducing the power of the statistic: This analysis is not based on 2005 citations to 2004 articles, plus 2006 citations to 2005 articles, plus 2007 citations to 2006 articles, etc. It is based on cumulative 2004-2008 citations to 2004, 2005, 2006 etc. articles, reckoned in June 2008. 2007 articles are not only younger: they are also more recent. Hence it is not clear what the Age/OA interaction in Table S2 really means: Has (1) the OA advantage for articles really been shrinking across those 4 years, or are citation rates for younger articles simply noisier, because based on smaller citation spans, hence (2) the OA Advantage *grows more detectable* as articles get older?)

From what is described and depicted in Figure 1, the natural interpretation of the Age/OA interaction seems to be the latter: As we move from one-year-old articles (2007) toward four-year-old articles, three things are happening: non-OA citations are growing with time, OA citations are growing with time, and the OA/non-OA Advantage is emerging with time.

> ***"[To] calculate… the estimated cost per citation [$400 - $9000]… we multiply the open access citation advantage for each journal (a multiplicative effect) by the impact factor of the journal… Considering [the] strong evidence of a decline of the citation advantage over time, the cost…would be much higher…"***

Although these costs are probably overestimated (because the OA Advantage is underestimated, and there is no decline but rather an increase) the thrust of these figures is reasonable: It is not worth paying for AC-OA for the sake of the AC-OA Advantage: It makes far more sense to get the OA Advantage for free, through OA Self-Archiving.

Note, however, that the potentially informative journal impact factor (JIF) was omitted from the full-model multiple regression equation across journals (#6). It should be tested. So should the percentage OA for each journal/year. And after that the analysis should be redone separately for, say, the four successive JIF quartiles. If adding the JIF to the equation reduces the OA Advantage further, whereas without JIF the OA Advantage increases in each successive quartile, then that implies that a big factor in the OA Advantage is the Quality Advantage (QA).

> ***"that we were able to explain some of the citation advantage by controlling for differences in article characteristics… strengthens the evidence that self-selection – not access – is the explanation for the citation advantage… more citable articles have a higher probability of being made freely accessible"***

Self-selection (QB) is undoubtedly one of the factors in the OA Advantage, but this analysis has not estimated the size of its contribution, relative to many other factors (AA, CA, DA, EA, QA). It has simply shown that some of *the same factors that influence citation counts, influence the OA citation Advantage too*.

By failing to test and control for the Quality Advantage in particular (by not testing JIFs in the full regression equation, by not taking percentage OA per journal/year into account, by restricting the sample-size for the highest impact, largest-sample journal, PNAS, by overlooking OA self-archiving and crediting it to non-OA, by not testing citation-brackets of JIF quartiles), the article needlessly misses the opportunity to analyze the factors contributing to the OA Advantage far more rigorously.

> ***“earlier studies [on the OA Advantage] may be showing an early-adopter effect…”***

This is probably true. And early adopters also have a Competitive Advantage (CA). But with only about 20% OA being provided overall today, the CA is still there, except if it can be demonstrated – as Davis certainly has not demonstrated – that the c. 20% of articles that are being made OA today correspond sufficiently closely to that top 20% of articles that receive 80% of all citations. (*Then the OA Advantage would indeed be largely QB*.)

> ***“authors who deposited their manuscripts in the arXiv tended to be more highly-cited than those who did not”***

There is some circularity in this, but it is correct to say that this correlation is compatible with both QB and QA, and probably both are contributing factors. But none of the prior studies nor this one actually estimate their relative contributions (nor those of AA, CA, DA and EA).

> ***“any relative citation advantage that was enjoyed by early adopters would disappear over time”***

It is not that CA (Competitive Advantage) disappears simply because time elapses: CA only disappears if the competitors provide OA too! The same is true of QB (Quality Bias), which also disappears once everyone is providing OA. But at 20%, we are nowhere near 100% OA yet; hence there is still plenty of scope for a competitive edge.

> ***“If a citation advantage is the key motivation of authors to pay open access fees, then the cost/benefit of this decision can be quite expensive for some journals.”***

This is certainly true, and would be true even if the OA citation Advantage were astronomically big – but the reason it is true is that *authors need not pay AC-OA fees for OA at all*: they can self-archive for free (and indeed are being increasingly mandated by their funders and institutions to do so).

> ***“Randomized controlled trials provide a more rigorous methodology for measuring the effect of access independently of other confounding***

> ***effects (Davis et al., 2008)… the differences we report in our study… have more likely explained the effect of self-selection (or self-promotion) than of open access per se."***

The syntax here makes it a little difficult to interpret, but if what is meant is that Davis et al's prior study has shown that the OA Advantage found in the present study was more likely to be a result of QB than of QA, AA, CA, DA, or EA, then it has to be replied that that prior study showed nothing of the sort (Harnad 2008b). All it showed was that one cannot detect a significant OA Advantage at all one year after publication when OA is randomized. (The same is true when OA is not randomized.)

However, the prior Davis et al. study did find a significant DA (Download Advantage) for OA articles in the first year. And other studies have reported a significant correlation between early downloads and later citations (Brody et al. 2006).

So the prior Davis et al. study (1) confirmed the familiar failure to detect the OA Advantage in the first year, and (2) found a significant DA in the first year (probably predictive of a later OA citation Advantage). The present Davis study found (i) a significant OA Advantage, (ii) smallest in the first year (2007), much bigger by the fourth (2004).

> ***"Retrospective analysis… our analysis is based on cumulative citations to articles taken at one point in time. Had we tracked the performance of our articles over time – a prospective approach – we would have stronger evidence to bolster our claim that the citation advantage is in decline. Still, we feel that cumulative citation data provides us with adequate inference."***

Actually, it would be possible, with a fuller analysis using the ISI database, to calculate not only the citation counts for each *cited* article, but the dates of the *citing* articles. So a "prospective" analysis can be done in retrospect. Without performing that analysis, however, the present study does not provide evidence of a decline in the OA Advantage with time, just evidence of an improved signal/noise ratio for measuring the OA Advantage with time. A "prospective" analysis, taking citing dates as well as cited dates into account, would be welcome (and is far more likely to show that the size of the OA Advantage is, if anything, growing, rather than confirming the author's interpretation, unwarranted on the present data, that it is shrinking).

> ***"all of the journals under investigation make their articles freely available after an initial period of time [hence] any [OA Advantage] would be during these initial months in which there exists an access differential between open access and subscription-access articles. We would expect therefore that the effect of open access would therefore be strongest in the earlier years of the life of the article and decline over time. In other words, we would expect our trend (Figure 1) to operate in the reverse direction."***

The reasoning here is a bit hard to follow, but the Kurtz studies that Davis cites show that in Astronomy, making preprints OA in the year or so before publication (after which all Astronomy postprints are OA) results in both "a strong EA effect and a strong [QB] effect." But even in a fast-moving field like Astronomy, the effect is not immediate! There is no way to predict from the data for Astronomy how quickly an EA effect for nonsubscribers during the embargo year in Biomedicine should make itself felt in citations, but it is a safe bet that, as with citation latency itself, and the latency of the OA citation Advantage, the "EmA" ("Embargo Access") counterpart of the EA effect in access-embargoed Biomedical journals will need a latency of a few years to become detectable. And since Davis's age/OA interaction, based on static, cumulative, retrospective data, is just as readily interpretable as indicating that OA Advantages require time and sample-size growth in order to occur and be detected, the two patterns are perfectly compatible.

> ***"we are at a loss to come up with alternative explanations to explain the monotonic decline in the citation advantage"***

*There is no monotonic decline to explain.* Just (a) low power in initial years, (b) cumulative data not analyzed to equate citing/cited year spans, (c) the failure to test for QA citation-bracket effects, and (d) the failure to reckon self-archiving OA into the OA Advantage (treating it instead as non-OA).

If this had been a JASIST referee report, I would have recommended performing several further analyses taking into account:

(1) self-archiving OA
(2) percentage OA per journal per year
(3) JIFs and citation-brackets
(4) the full PNAS dataset
(5) citing-article-date vs cited-article-date

and making the interpretation of the resultant findings more even-handed, rather than slanting toward the author's preferred hypothesis that the OA Advantage is due solely or mostly to QB.

> **Full Disclosure:** *I am an OA advocate. And although I hope that I do not have a selective confirmation bias favoring QA, AA, CA, DA & EA, and against the Quality Bias (QB), I do think it is particularly important to ensure that QB is not given more weight than it has been empirically demonstrated to be able to bear.*
> *Davis writes:*
>
> > "Free dissemination of the scientific literature may [1] speed up the transfer of knowledge to industry, [2] enable scientists in poor and developing countries to access more information, and [3] empower the general public. There are clearly many benefits to making one's research findings freely available to the general public – a citation advantage may not be one of them."
>
> *It may not be; but if it is not, then it must be demonstrated rigorously that it is not, both because of the prima facie evidence from download and citations, and because of what is at stake:*

*Currently, most researchers (c. 85%) are not yet providing OA to their research (Björk et al. 2008). Benefits [1]-[3] have clearly proven insufficient to motivate researchers to provide OA: only a small proportion of research articles have industrial applications [1] or general-public appeal [3], and most researchers are alas not motivated by charity to poor and developing countries [2]). Hence motivation to provide OA depends on other expected benefits.*

*Kurtz has suggested that QB itself may provide the motivation: seeing that the better research is being made OA by the better researchers should motivate all researchers to do likewise.*

*But on the assumption that the better researchers are making their better research OA for a reason, rather than just [1]-[3] (or just to imitate one another), it seems reasonable to ask what benefits their motivation is based on. And the enhancement of the uptake, usage, application and impact of their own research seems to be the obvious candidate. Otherwise, providing OA just for its own sake seems an irrational practice, especially on the part of the better researchers, for the better research.*

*A rigorous estimate of the true causal contribution of QB to the observed OA Advantage is also important for research policy reasons:*

*Although they have been cited in support of OA mandates by institutions and funders, industrial applications [1], charity [2] and public appeal [3] are not in themselves sufficient to motivate the adoption of such mandates, or compliance with them. Research impact itself would seem to be the natural rationale for maximizing research access.*

*So if the observed OA Advantage really does not provide the benefits that would motivate researchers to provide OA, this has to be demonstrated directly and rigorously. Why would the better researchers make their better research OA if it conferred no benefits (other than [1]-[3])? Too much is at stake to accept QB as the default hypothesis. The natural default hypothesis on the basis of the prima facie evidence (more downloads and citations) is that increasing the access to research increases the usage and impact of research. The higher download and citation counts correlated with OA support this. The burden of proof hence falls on providing evidence to the contrary.*